\documentclass[12pt]{article}

\usepackage{graphicx}

\def\gtwid{\mathrel{\raise.3ex\hbox{$>$\kern-.75em\lower1ex\hbox{$\sim$}}}}
\def\ltwid{\mathrel{\raise.3ex\hbox{$<$\kern-.75em\lower1ex\hbox{$\sim$}}}}
\def\square{\kern1pt\vbox{\hrule height 1.2pt\hbox{\vrule width 1.2pt\hskip 3pt
   \vbox{\vskip 6pt}\hskip 3pt\vrule width 0.6pt}\hrule height 0.6pt}\kern1pt}

\begin{document}

\begin{titlepage}

\begin{flushright}
UFIFT-QG-16-01 \\
CCTP-2016-04 \\
CCQCN-2016-135 \\
ITCP-IPP 2016/05
\end{flushright}

\vskip 1cm

\begin{center}
{\bf Improving the Single Scalar Consistency Relation}
\end{center}

\vskip .5cm

\begin{center}
D. J. Brooker$^{1*}$, N. C. Tsamis$^{2\star}$ and R. P. Woodard$^{1\dagger}$
\end{center}

\vskip .5cm

\begin{center}
\it{$^{1}$ Department of Physics, University of Florida,\\
Gainesville, FL 32611, UNITED STATES}
\end{center}

\begin{center}
\it{$^{2}$ Institute of Theoretical Physics \& Computational Physics, \\
Department of Physics, University of Crete, \\
GR-710 03 Heraklion, HELLAS}
\end{center}

\vspace{.5cm}

\begin{center}
ABSTRACT
\end{center}
We propose a test of single-scalar inflation based on using the well-measured
scalar power spectrum to reconstruct the tensor power spectrum, up to a single
integration constant. Our test is a sort of integrated version of the single-scalar
consistency relation. This sort of test can be used effectively, even when the
tensor power spectrum is measured too poorly to resolve the tensor spectral
index. We give an example using simulated data based on a hypothetical detection 
with tensor-to-scalar ratio $r = 0.01$. Our test can also be employed for 
correlating scalar and tensor features in the far future when the data is good.

\begin{flushleft}
\end{flushleft}

\vskip .5cm

\begin{flushleft}
$^{*}$ e-mail: djbrooker@ufl.edu \\
$^{\star}$ e-mail: tsamis@physics.uoc.gr \\
$^{\dagger}$ e-mail: woodard@phys.ufl.edu
\end{flushleft}

\end{titlepage}

\section{Introduction}

The theory of primordial inflation \cite{Brout:1977ix,Starobinsky:1980te,
Kazanas:1980tx,Sato:1980yn,Guth:1980zm,Linde:1981mu,Albrecht:1982wi,
Linde:1983gd} has had a profound effect on cosmology and fundamental theory.
Particularly striking is the prediction that primordial tensor 
\cite{Starobinsky:1979ty} and scalar \cite{Mukhanov:1981xt} perturbations 
derive from quantum gravitational fluctuations which fossilized near 
the end of inflation. This not not only affords us access to quantum gravity 
at an intoxicating energy scale \cite{Woodard:2009ns,Ashoorioon:2012kh,
Krauss:2013pha}, it also provides information about the mechanism that 
powered inflation. This information can be accessed by comparing observations 
of the two power spectra, $\Delta^2_{\mathcal{R}}(k)$ and $\Delta^2_{h}(k)$, 
to predictions from the many models \cite{Mukhanov:1990me,Liddle:1993fq,
Lidsey:1995np}. For example, the simplest models of inflation are driven by 
the potential of a single, minimally coupled scalar. These models all obey 
the single-scalar consistency relation \cite{Polarski:1995zn,GarciaBellido:1995fz,
Sasaki:1995aw},
\begin{equation}
r \approx -8 n_t \; , \label{single}
\end{equation}
where $r$ is the tensor-to-scalar ratio and $n_t$ is the tensor spectral
index,
\begin{equation}
r(k) \equiv \frac{\Delta^2_{h}(k)}{\Delta^2_{\mathcal{R}}(k)} \qquad ,
\qquad n_t(k) \equiv \frac{\partial \ln(\Delta^2_{h}(k))}{\partial \ln(k)}
\; . \label{rntdef}
\end{equation}
A statistically significant violation of (\ref{single}) would falsify the
entire class of single-scalar models, as well as all models which are related
to them by conformal transformation, such as $f(R)$ inflation 
\cite{Brooker:2016oqa}.

Although the single-scalar consistency relation was a brilliant theoretical
insight, the progress of observation has rendered it somewhat inconvenient.
The scalar power spectrum was first resolved in 1992 \cite{Smoot:1992td},
and is now quite well measured \cite{Hinshaw:2012aka,Hou:2012xq,
Sievers:2013ica,Ade:2013zuv}. The tensor power spectrum has not yet been
resolved \cite{Adam:2014bub,Ade:2015tva}, but polarization measurements 
are now providing the strongest limits on it \cite{Ade:2015xua}. It is not
known if the current generation of polarization experiments 
\cite{Hattori:2013jda,Lazear:2014bga,Rahlin:2014rja,Ahmed:2014ixy,
MacDermid:2014wca} can resolve the tensor power spectrum at all, and it is 
very unlikely that they will measure it well enough to constrain the tensor
spectral index with much accuracy.

In view of the observational situation, it makes sense to develop a test of 
single-scalar inflation that is based primarily on the abundant data for 
$\Delta^2_{\mathcal{R}}(k)$, and does not require taking derivatives of
the sparse data for $\Delta^2_{h}(k)$ likely to result from the first 
positive detections. There is no reason not to do this because the close 
relation between the tensor and scalar mode functions of single-scalar 
inflation implies that either power spectrum determines the other, up to 
some integration constants. That is the purpose of this paper. In the next 
section we fix notation, recall the relation between the two power spectra, 
and infer the tensor power spectrum from the scalar one. Section 3 gives a
comparison between the single scalar consistency relation and the scatter
test we propose, using simulated data based on a hypothetical detection of
$r = 0.01$ with the same number of data points and the same fractional
error as was in fact reported by the recent spurious BICEP2 detection
\cite{Ade:2014xna}. The final section mentions applications.

\section{Constructing $\Delta^2_{h}(k)$ from $\Delta^2_{\mathcal{R}}(k)$}

We work in spatially flat, co-moving coordinates with scale factor $a(t)$,
Hubble parameter $H(t)$ and first slow roll parameter $\epsilon(t)$,
\begin{equation}
ds^2 = -dt^2 + a^2(t) d\vec{x} \!\cdot\! d\vec{x} \quad \Longrightarrow \quad
H(t) \equiv \frac{\dot{a}}{a} \quad , \quad \epsilon(t) \equiv -
\frac{\dot{H}}{H^2} \; . \label{geometry}
\end{equation}
We assume $a(t)$ is known, with the scalar background and potential
determined to enforce the background Einstein equations
\cite{Tsamis:1997rk,Saini:1999ba,Capozziello:2005mj,Woodard:2006nt,Guo:2006ab},
\begin{eqnarray}
\varphi_0(t) & = & \varphi_0(t_i) \pm \int_{t_i}^{t} \!\! dt' H(t')
\sqrt{ \frac{\epsilon(t')}{4 \pi G}} \quad \Longleftrightarrow \quad t(\varphi)
\; , \label{phirecon} \\
V(\varphi) & = & \frac{[3 \!-\! \epsilon(t)] H^2(t)}{8 \pi G} \Biggl\vert_{
t=t(\varphi)} \; . \label{Vrecon}
\end{eqnarray}
We fix the gauge so that the full scalar agrees with its background value and
the graviton field $h_{ij}$ is transverse, with $g_{00}$ and $g_{0i}$ regarded
as constraints. The two dynamical fields are $h_{ij}$ and $\zeta$, which reside
in the 3-metric $g_{ij} = a^2 e^{2\zeta} [e^h]_{ij}$. At quadratic order their 
Lagrangian is \cite{Woodard:2014jba},
\begin{equation}
\mathcal{L}_2 = \frac{a^3}{64 \pi G} \Bigl[ \dot{h}_{ij} \dot{h}_{ij} -
\frac{h_{ij,k} h_{ij ,k}}{a^2}\Bigr] + \frac{\epsilon a^3}{8\pi G} \Bigl[
\dot{\zeta}^2 - \frac{\zeta_{,k} \zeta_{,k}}{a^2} \Bigr] \; . \label{Lag2}
\end{equation}
The spatial plane wave mode functions of the graviton are $u(t,k)$, with exactly
the same polarization tensors as in flat space. From (\ref{Lag2}) we see that the
evolution equation, Wronskian and asymptotically early form of the tensor mode 
functions $u(t,k)$ are,
\begin{equation}
\ddot{u} + 3 H \dot{u} + \frac{k^2}{a^2} u = 0 \; , \; u \dot{u}^* \!-\! 
\dot{u} u^* = \frac{i}{a^3} \; , \; u(t,k) \longrightarrow 
\frac{\exp[-ik \int_{t_i}^t \frac{dt'}{a(t')}]}{\sqrt{2 k a^2(t)}} \; .
\label{ueqns}
\end{equation}
The scalar perturbation $\zeta$ has spatial plane wave mode functions $v(t,k)$.
From (\ref{Lag2}) we see that their evolution equation, Wronskian and 
asymptotically early form are,
\begin{equation}
\ddot{v} + \Bigl(3 H + \frac{\dot{\epsilon}}{\epsilon}\Bigr) \dot{v} + 
\frac{k^2}{a^2} v = 0 \; , \; v \dot{v}^* \!- \dot{v} v^* = \frac{i}{\epsilon a^3} 
\; , \; v(t,k) \longrightarrow \frac{\exp[-ik \int_{t_i}^t \frac{dt'}{a(t')}]}{
\sqrt{2 k \epsilon(t) a^2(t)}}  \; . \label{veqns}
\end{equation}
The two power spectra are determined (at tree order) by evolving their
respective mode functions from their early forms through the time $t_k$ of 
first horizon crossing ($k \equiv H(t_k) a(t_k)$), after which they approach
constants,
\begin{eqnarray}
\Delta^2_{\mathcal{R}}(k) & = & \frac{k^3}{2 \pi^2} \times 4 \pi G
\times \Bigl\vert v(t,k)\Bigr\vert^2_{t \gg t_k} \approx 
\frac{G H^2(t_k)}{\pi \epsilon(t_k)} \; , \label{DR} \\
\Delta^2_{h}(k) & = & \frac{k^3}{2\pi^2} \times 32\pi G \times
2 \times \Bigl\vert u(t,k)\Bigr\vert^2_{t \gg t_k} 
\approx \frac{16 G H^2(t_k)}{\pi} \; . \label{Dh}
\end{eqnarray}

The relations (\ref{ueqns}) which define $u(t,k)$ are carried into the
relations (\ref{veqns}) which define $v(t,k)$ by making simultaneous
changes in the scale factor and the co-moving time \cite{Tsamis:2003zs,
Romania:2012tb},
\begin{equation}
a(t) \longrightarrow \sqrt{\epsilon(t)} \, a(t) \qquad , \qquad
\frac{\partial}{\partial t} \longrightarrow \frac1{\sqrt{\epsilon(t)}}
\, \frac{\partial}{\partial t} \; . \label{trans}
\end{equation}
To understand what this means for the power spectra we must consider them
as nonlocal functionals of the expansion history $a(t)$, which will involve
integrals and derivatives with respect to time. We denote this functional
dependence with square brackets, so relation (\ref{trans}) implies,
\begin{equation}
\Delta^2_{\mathcal{R}}\Bigl[a,dt\Bigr](k) = \frac1{16} \Delta^2_h\Bigl[
\sqrt{\epsilon} a,\sqrt{\epsilon} dt\Bigr](k) \; . \label{Dtrans}
\end{equation}
Relation (\ref{Dtrans}) is easy to check at leading slow roll order by 
comparing the slow roll approximation for $\Delta^2_{\mathcal{R}}(k)$ on 
the right hand side of (\ref{DR}) with the effect of making transformation 
(\ref{trans}) on the Hubble parameter in the right hand side of 
expression (\ref{Dh}),
\begin{equation}
H(t) \equiv \frac{\partial}{\partial t} \, \ln[a(t)]
\longrightarrow \frac1{\sqrt{\epsilon}} \frac{\partial}{\partial t}
\, \ln\Bigl[ \sqrt{\epsilon} \, a\Bigr] = \frac{H \!+\! 
\frac{\dot{\epsilon}}{2 \epsilon}}{\sqrt{\epsilon}} \; .
\end{equation}
However, we stress that relation (\ref{Dtrans}) is exact, not just valid 
at leading slow roll order, provided one employs the exact expressions 
for $\Delta^2_{h}(k)$ and $\Delta^2_{\mathcal{R}}(k)$. 

\begin{figure}[ht]
\includegraphics[width=6.0cm,height=4.8cm]{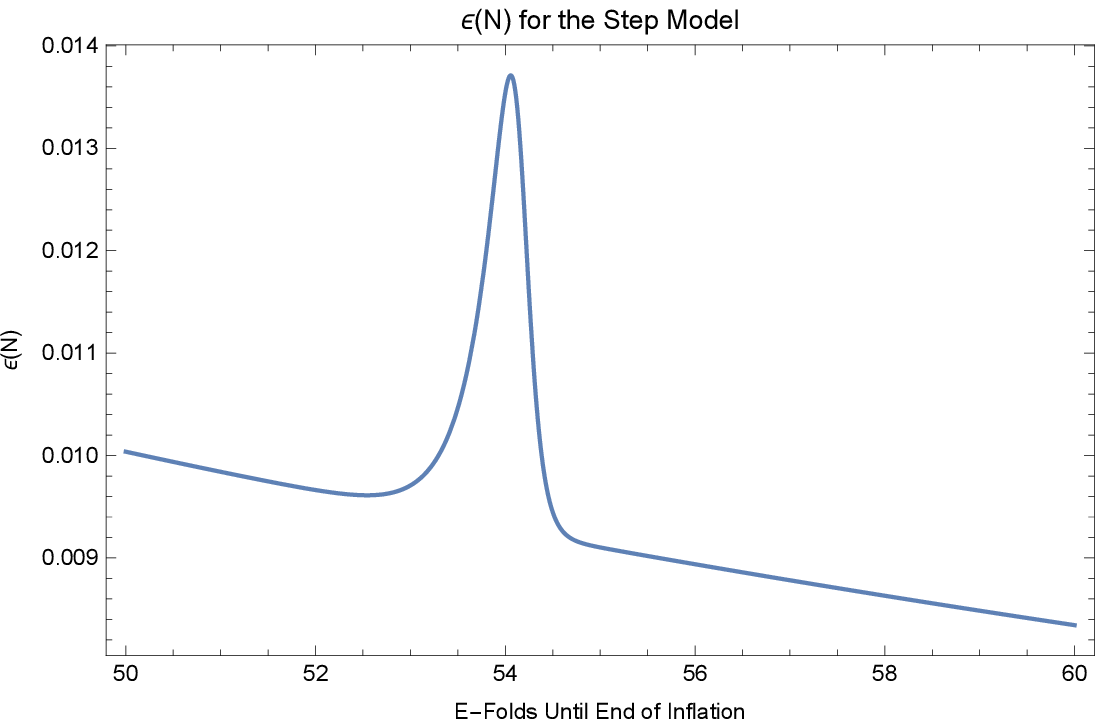}\hskip 1cm
\includegraphics[width=6.0cm,height=4.8cm]{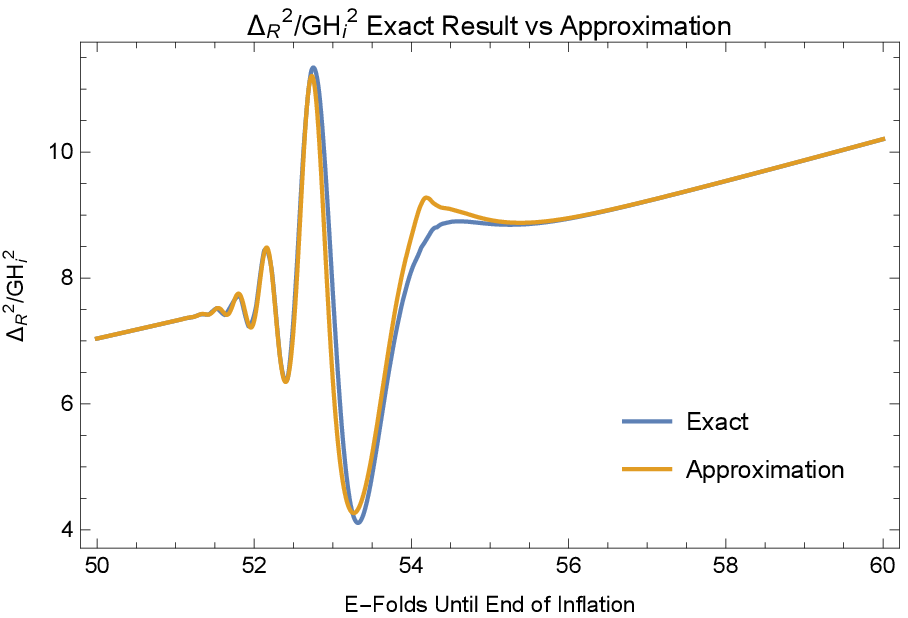}
\caption{The left hand figure shows the first slow roll parameter for a 
model which was proposed \cite{Adams:2001vc,Mortonson:2009qv} to explain 
the observed features in the scalar power spectrum at $\ell \approx 22$ 
and $\ell \approx 40$ which are visible in the data reported from both  
WMAP \cite{Covi:2006ci,Hamann:2007pa} and PLANCK \cite{Hazra:2014goa,
Hazra:2016fkm}. The right hand figure shows the resulting scalar 
power spectrum (in blue), with the result of our analytic approximation 
(\ref{scalarapprox}) (in yellow). The slow roll approximation (\ref{DR}) 
does not give a very accurate fit even to the main feature in the range 
$54.5 < N < 53$ e-foldings before the end of inflation, and it completely
misses the secondary oscillations visible in the range $53.5 < N < 51.5$.
The nonlocal contributions (\ref{scalarnonlocal}) are essential for
correctly reproducing these features.}
\label{approximation}
\end{figure}

We should also point out that very accurate functional expressions are 
now available for the power spectra of single scalar inflation, valid to 
all orders in the slow roll parameter $\epsilon(t_k)$, and even including 
nonlocal effects from times before and after first horizon crossing 
\cite{Brooker:2015iya,Brooker:2016xkx}. These expressions take the form 
\cite{Brooker:2017kjd},
\begin{eqnarray}
\Delta^2_{\mathcal{R}}(k) & \simeq & \frac{G H^2(t_k)}{\pi \epsilon(t_k)} 
\!\times\! C\Bigl(\epsilon(t_k)\Bigr) \!\times\! \exp\Bigl[
\sigma[\epsilon](k)\Bigr] \; , \label{scalarapprox} \\
\Delta^2_{h}(k) & \simeq & \frac{16 G H^2(t_k)}{\pi} \!\times\! C\Bigl(
\epsilon(t_k)\Bigr) \!\times\! \exp\Bigl[\tau[\epsilon](k)\Bigr] \; , 
\label{tensorapprox}
\end{eqnarray}
where the local slow roll correction factor is,
\begin{equation}
C(\epsilon) \equiv \frac1{\pi} \Gamma^2\Bigl( \frac12 \!+\! 
\frac1{1 \!-\! \epsilon}\Bigr) \Bigl[ 2 (1 \!-\! \epsilon)
\Bigr]^{\frac2{1-\epsilon}} \approx 1 - \epsilon \; . \label{Cdef}
\end{equation}
For the nonlocal corrections $\sigma[\epsilon](k)$ and $\tau[\epsilon](k)$ 
it is best to abuse the notation by writing the first slow parameter 
$\epsilon(n) \equiv \epsilon(t(n))$ as a function of $n \equiv \ln[a(t)/a_i]$, 
the number of e-foldings since the start of inflation,
\begin{eqnarray}
\sigma[\epsilon](k) & = & \int_{0}^{n_k} \!\!\!\!\! dn \Biggl[
\partial_n^2 \ln[\epsilon(n)] \!+\! \frac12 \Bigl( \partial_n 
\ln[\epsilon(n)]\Bigr)^2 \!+\! 3 \partial_n \ln[ \epsilon(n)]
\Biggr] G\Bigl(e^{\Delta n}\Bigr) \nonumber \\
& & \hspace{2cm} - \partial_{n_k} \ln[\epsilon(n_k)] \, G(1) +
\int_{n_k}^{\infty} \!\!\!\!\! dn \, \partial_n \ln[\epsilon(n)]
\frac{2 G(e^{\Delta n})}{1 \!+\! e^{2 \Delta n}} \; , 
\label{scalarnonlocal} \qquad \\
\tau[\epsilon](k) & = & \int_0^{n_k} \!\!\!\!\! dn \Biggl[ \mathcal{E}_1\Bigl(
e^{\Delta n}\Bigr) \epsilon''(n) \!+\! \mathcal{E}_2\Bigl( e^{\Delta n}\Bigr)
\Bigl(\epsilon'(n)\Bigr)^2 \!\!+\! \mathcal{E}_3\Bigl(e^{\Delta n}\Bigr)
\epsilon'(n)\Biggr] G\Bigl(e^{\Delta n}\Bigr) \nonumber \\
& & \hspace{-2cm} - \epsilon'(n_k) \mathcal{E}_1(1) G(1) -\!\! \int_{n_k}^{\infty}
\!\!\!\!\!\!\! dn \Biggl\{ \Delta \epsilon(n) \!+\! \Bigl(\frac{4 \!+\! 2 
e^{2 \Delta n}}{1 \!+\! e^{2 \Delta n}}\Bigr) \!\! \int_{n_k}^{n} \!\!\!\!\! dm 
\, \Delta \epsilon(m) \!\Biggr\} \frac{2 G(e^{\Delta n})}{1 \!+\! e^{2 \Delta n}} 
\; . \label{tensornonlocal}
\end{eqnarray}
Here $\Delta n \equiv n - n_k$, $\Delta \epsilon(m) \equiv \epsilon(m) - \epsilon_k$,
and the functions of $x \equiv e^{\Delta n}$ are,
\begin{eqnarray}
G(x) & = & \frac12 \Bigl(x \!+\! x^3\Bigr) \sin\Bigl[\frac{2}{x} \!-\! 2
{\rm arctan}\Bigl(\frac1{x}\Bigr)\Bigr] \; , \\
\mathcal{E}_1(x) & \simeq & = \frac{\frac12 x^2 \!-\! 1.8 x^4 \!+\! 1.5 x^6 
\!-\! 0.63 x^8}{1 \!+\! x^2} \; , \\ 
\mathcal{E}_2(x) & \simeq & \frac{2.8 x^4 \!-\! 7 x^6 \!+\! 3 x^8 \!+\! 1.8 
x^{10} \!-\! 2.3 x^{12} \!+\! 0.95 x^{14} \!-\! 0.20 x^{16}}{(1 \!+\! x^2)^2}
\; , \qquad \\ 
\mathcal{E}_3(x) & \simeq & \frac{\frac92 x^2 \!-\! 11.9 x^4 \!+\! 7.1 x^6 \!-\! 
1.3 x^8 \!-\! 1.9 x^{10}}{(1 \!+\! x^2)^2} \; .
\end{eqnarray}
The $95\%$ confidence bound on the tensor-to-scalar ratio of $r < 0.12$ \cite{Ade:2015tva,Ade:2015xua} implies $\epsilon < 0.0075$, so $\tau[\epsilon](k)$
is about a hundred times smaller than $\sigma[\epsilon](k)$. Models with smooth
potentials typically have $\epsilon' \sim \epsilon^2$ and $\epsilon'' \sim 
\epsilon^3$, so the leading contributions in $\sigma[\epsilon](k)$ come from the
3rd and 5th terms of expression (\ref{scalarnonlocal}). In particular the 5th
(final) term is needed to correct for a systematic under-prediction of the local
slow roll approximation \cite{Brooker:2017kjd}. For models with features the
leading contributions to $\sigma[\epsilon](k)$ come from the 1st, 3rd and 4th 
terms of expression (\ref{scalarnonlocal}) \cite{Brooker:2017kjd}. These 
corrections can be very important for realistic models such as the one depicted 
in Figure \ref{approximation}.

To keep the analysis simple, we illustrate the procedure for predicting 
$\Delta^2_h(k)$ from $\Delta^2_{\mathcal{R}}(k)$ using only the leading slow 
roll terms in expressions (\ref{scalarapprox}-\ref{tensorapprox}), without 
either of the nonlocal corrections or even the slow roll factor $C(\epsilon)$. 
The conversion from wave number to time is,
\begin{equation}
k = H(t_k) a(t_k) \qquad \Longrightarrow \qquad \frac{dk}{k} = (1 \!-\!
\epsilon) H dt \approx H dt \; . \label{1stcross}
\end{equation}
The leading slow roll approximation (\ref{scalarapprox}) for the scalar power 
spectrum can be recognized as a differential equation for the Hubble
parameter,
\begin{equation}
\Delta^2_{\mathcal{R}}(k) \simeq \frac{G H^2(t_k)}{\pi \epsilon(t_k)} 
= -\frac{G H^4(t_k)}{\pi \dot{H}(t_k)} \; .
\end{equation}
We can integrate this equation from some arbitrary time $t_*$ to $t_k$,
\begin{equation}
d\Bigl( \frac1{H^2}\Bigr) \simeq \frac{2 G \, d\ln(k)}{\pi \Delta^2_{
\mathcal{R}}(k)} \quad \Longrightarrow \quad \frac1{H^2(t_k)} - 
\frac1{H^2(t_*)} \simeq \frac{2 G}{\pi} \int_0^{\ln(k/k_*)} 
\frac{d\ln(k')}{\Delta^2_{\mathcal{R}}(k')} \; . \label{Hrecon}
\end{equation}
Substituting the reconstructed Hubble parameter (\ref{Hrecon}) into 
the leading slow roll approximation (\ref{Dh}) for the tensor power
spectrum gives,
\begin{equation}
\Delta^2_{h}(k) \simeq \frac{16 G H^2(t_k)}{\pi} \simeq 
\Delta^2_{h}(k_*) \Biggl[1 + \frac{r(k_*)}{8} \int_0^{\ln(k/k_*)} 
\!\!\!\!\!\!\!\!\!\!\!\!\!\!\! \, d\ln(k') 
\frac{\Delta^2_{\mathcal{R}}(k_*)}{\Delta^2_{\mathcal{R}}(k')} 
\Biggr]^{-1} \; . \label{result}
\end{equation}
Equation (\ref{result}) is in some sense an integrated form of the
single-scalar consistency relation (\ref{single}) which can be applied
more reliably. Both relations are valid to leading slow roll order, but 
whereas (\ref{single}) compares a single value of the high quality data 
in $\Delta^2_{\mathcal{R}}(k)$ with a derivative of the poor data on 
$\Delta^2_{h}(k)$, our relation (\ref{result}) combines a single 
measurement of the tensor power spectrum at $ k = k_*$ with the high 
quality scalar data to predict what $\Delta^2_{h}(k)$ should be for 
other wave numbers. This seems to be a better way of exploiting the 
sparse data on $\Delta^2_{h}(k)$ which is likely to persist for some 
years after a first positive detection.

\section{Comparison Using Simulated Data}

It is illuminating to compare the single scalar consistency relation
with the method we propose using simulated data. Let us suppose that the
actual tensor power spectrum corresponds to single scalar inflation 
with $r = \frac1{100}$, and which implies $n_t = -\frac1{800}$. We 
further suppose the simplest possible $k$ dependence,
\begin{equation}
\Delta_{h}^2(k) = r A_S \Bigl( \frac{k}{k_0}\Bigr)^{n_t} \qquad 
\Longrightarrow \qquad \ln\Bigl[\Delta^2_{h}(k)\Bigr] = \ln\Bigl[ r
A_S\Bigr] + n_t \!\times\! \ln\Bigl[\frac{k}{k_0}\Bigr] \; , \label{relation}
\end{equation}
where the scalar amplitude (at the tensor pivot $k_0$) is $A_S = 2.5 \times 
10^{-9}$. Let us assume that the first positive detection of this tensor
power spectrum consists of results for five binned wave numbers, the same as 
was in fact reported for the spurious BICEP2 detection \cite{Ade:2014xna}. 
To simplify matters we assume a linear relation for logarithms of the 
observed wave numbers, $\ln[k_{i+1}/k_{i}] = \frac13$, and that each 
measurement of $\ln[\Delta^2_{h}]$ has the same 1-sigma uncertainty of 
$\sigma = \frac14$. These numbers are roughly consistent with what BICEP2
actually reported \cite{Ade:2014xna}. Hence the detection consists of five
observations $y_i$ obeying the relation,
\begin{equation}
y_i = \ln\Bigl[2.5 \!\times\ 10^{-11}\Bigr] - \frac{i}{2400} + e_i \qquad
, \qquad i \in \Bigl\{1,2,3,4,5\Bigr\} \; , \label{simulation}
\end{equation}
where the $e_i$ are independent Gaussian random variables with mean zero and 
standard deviation $\sigma = \frac14$. Table \ref{data} simulates the five 
data points using a random number generator to find the $e_i$.

\begin{table}
\setlength{\tabcolsep}{8pt}
\def\arraystretch{1.5}
\centering
\begin{tabular}{|@{\hskip 1mm }c@{\hskip 1mm }||c|c|c|c|}
\hline
$i$ & $\ln(2.5 \times 10^{-11})$ & $-\frac{i}{2400}$ & $e_i$ & $y_i$ \\
\hline\hline
1 & $-24.412145$ & $-0.000417$ & $+0.226742$ & $-24.185820$ \\
\hline
2 & $-24.412145$ & $-0.000833$ & $-0.176041$ & $-24.589020$ \\
\hline
3 & $-24.412145$ & $-0.001250$ & $-0.091555$ & $-24.504950$ \\
\hline
4 & $-24.412145$ & $-0.001667$ & $-0.164330$ & $-24.578142$ \\
\hline
5 & $-24.412145$ & $-0.002083$ & $+0.331640$ & $-24.082589$ \\
\hline
\end{tabular}
\caption{Simulated data from relation (\ref{simulation}), representing
a hypothetical first detection of the tensor power spectrum with
$r = \frac1{100}$ and $n_t = -\frac1{800}$. The random errors $e_i$
follow a normal distribution with mean zero and standard deviation
$\sigma = \frac14$. \label{data}}
\end{table}

Because the relation (\ref{relation}) is linear we can use least squares
to determine the parameters. The least squares fit for $N$ data points
obeying the relation $y_i = \alpha + \beta x_i$ (with $x_i = i/3$) is,
\begin{eqnarray}
\alpha & = & \frac{\sum_{i=1}^{N} x_i^2 \sum_{j=1}^{N} y_j \!-\! 
\sum_{i=1}^{N} x_i \sum_{j=1}^{N} x_j y_j}{N \sum_{i=1}^{N} x_i^2
\!-\! (\sum_{i=1}^{N} x_i)^2} = \frac{ \sum_{i=1}^{N} \sum_{j=1}^{N} 
x_i (x_i \!-\! x_j) y_j}{\sum_{i=2}^{N} \sum_{j=1}^{i-1} (x_i \!-\! x_j)^2} 
\; , \qquad \label{alpha1} \\
\beta & = &  \frac{ N \sum_{i=1}^{N} x_i y_i - \sum_{i=1}^{N} x_i 
\sum_{j=1}^{N} y_j}{N \sum_{i=1}^{N} x_i^2 - (\sum_{i=1}^{N} x_i)^2}
= \frac{ \sum_{i=2}^{N} \sum_{j=1}^{i-1} (x_i \!-\! x_j) (y_i \!-\! y_j)}{
\sum_{i=2}^{N} \sum_{j=1}^{i-1} (x_i \!-\! x_j)^2} \; . \qquad \label{beta1}
\end{eqnarray}
Even in this general form it is obvious that expression (\ref{alpha1}) for
$\alpha$ represents a sort of average whereas expression (\ref{beta1}) is a
kind of numerical derivative. So we expect the fractional error on $\beta$ to
be larger than that on $\alpha$. That becomes even more apparent when 
specializing to $N = 5$ and $x_i = i/3$,
\begin{eqnarray}
\alpha & \longrightarrow & \frac{(8 y_1 \!+\! 5 y_2 \!+\! 2 y_3 \!-\! y_4 \!-\!
4 y_5)}{10} \simeq -24.453306 \pm 0.262202 \; , \label{alpha2} \\
\beta & \longrightarrow & \frac{(-6 y_1 \!-\! 3 y_2 \!+\! 3 y_4 \!+\! 6 y_5)}{10}
\simeq +0.065202 \pm 0.237171 \; . \label{beta2} 
\end{eqnarray}   
Hence the simulated data of Table \ref{data} implies a reasonably accurate 
reconstruction of the tensor-to-scalar ratio,
\begin{equation}
r = \exp\Bigl[ \alpha - \ln\Bigl(2.5 \!\times\! 10^{-9}\Bigr) \Bigr] = 0.0096
\pm 0.0027 \; , \label{reconr}
\end{equation}
but a miserably inaccurate value for the tensor spectral index,
\begin{equation}
n_t = \beta = 0.065 \pm 0.237 \; . \label{reconnt}
\end{equation}
The resulting check of the single scalar consistency relation is not very
sensitive,
\begin{equation}
0.010 \pm 0.003 = -0.522 \pm 1.897 \; . \label{badcheck}
\end{equation}
Because of the large (but statistically allowed) positive fluctuation $e_5$ 
the measured tensor spectral index (\ref{reconnt}) does not even have the 
right sign!

\begin{table}
\setlength{\tabcolsep}{8pt}
\def\arraystretch{1.5}
\centering
\begin{tabular}{|@{\hskip 1mm }c@{\hskip 1mm }||c|c|c|c|}
\hline
$i$ & $\alpha$ & $-\frac{r}{24} \times i$ & $z_i$ & $y_i - z_i$ \\
\hline\hline
1 & $-24.453306$ & $-0.000400$ & $-24.453706$ & $+0.267886$ \\
\hline
2 & $-24.453306$ & $-0.000800$ & $-24.454106$ & $-0.134914$ \\
\hline
3 & $-24.453306$ & $-0.001200$ & $-24.454506$ & $-0.050445$ \\
\hline
4 & $-24.453306$ & $-0.001599$ & $-24.454906$ & $-0.123236$ \\
\hline
5 & $-24.453306$ & $-0.001999$ & $-24.455306$ & $+0.372717$ \\
\hline
\end{tabular}
\caption{Predicted results according to relation (\ref{predictions}),
with the parameters $\alpha$ and $r$ taken from expressions
(\ref{alpha2}) and (\ref{reconr}), respectively. \label{scatter}}
\end{table}

We propose to instead use the much better measured scalar spectral
index to predict the tensor spectral index, up to an integration
constant, and then to compare the fluctuation of the observed data
around this prediction. For the model in question this might amount
to assuming predictions of the form,
\begin{equation}
z_i = \alpha - \frac{r}{24} \!\times \! i \; , \label{predictions}
\end{equation}
where $\alpha$ is (\ref{alpha2}) and $r$ is (\ref{reconr}). Table 
\ref{scatter} reports these predictions, along with the difference 
between each simulated observation $y_i$ and the associated prediction 
$z_i$. Of course the parameter $r$ comes from the parameter $\alpha$
through relation (\ref{reconr}), so the final column of Table \ref{scatter} 
represents four statistically independent measurements. The resulting 
estimate for the scatter between measurement and prediction is,
\begin{equation}
\sqrt{\frac14 \sum_{i=1}^{5} (y_i \!-\! z_i)^2} \simeq 0.246614 \; .
\end{equation}
This is quite consistent with our assumed 1-sigma fluctuation of 
$\sigma = \frac14$ for each observation. 

\section{Discussion}

Resolving the tensor power spectrum $\Delta^2_{h}(k)$ is crucial for 
the progress of inflation because it constrains the causative mechanism. 
This is already evident from the angst \cite{Ijjas:2013vea,Guth:2013sya,
Linde:2014nna,Ijjas:2014nta} elicited by the increasingly tight bounds 
on the tensor-to-scalar ratio $r$ \cite{Ade:2015lrj}. A positive detection 
at several different wave lengths has the potential to falsify entire 
classes of models. For example, any model in which inflation is driven 
by the potential of a minimally coupled scalar must obey relation 
(\ref{single}) between $r$ and the tensor spectral index $n_t$ 
\cite{Polarski:1995zn,GarciaBellido:1995fz,Sasaki:1995aw}. Unfortunately, 
relation (\ref{single}) requires taking a derivative of $\Delta^2_{h}(k)$, 
and the first generation of detections will probably be too sparse to 
provide a good bound because numerical differentiation makes bad data 
worse.

It makes more sense to integrate the high quality data we already possess 
for $\Delta^2_{\mathcal{R}}(k)$. If the leading slow roll expressions 
(\ref{DR}-\ref{Dh}) are assumed then the prediction (\ref{result}) from 
$\Delta^2_{\mathcal{R}}(k)$ requires only a single integration constant 
from $\Delta^2_{h}(k)$. (The same thing would be true even if the more 
accurate approximations (\ref{scalarapprox}-\ref{tensorapprox}) were 
employed \cite{Brooker:2017kjd}.) Fixing this constant uses up one 
combination of whatever data we have for $\Delta^2_{h}(k)$, leaving the 
scatter of the remaining data about the prediction as a legitimate test 
of single scalar inflation. Hence relation (\ref{result}) is a sort of
integrated form of the single-scalar consistency relation (\ref{single}) 
which can be applied more reliably. Section 3 compares this sort of scatter 
test with checking $r = -8 n_t$ for simulated data based on a hypothetical 
detection of $r = 0.01$ at five wave lengths with fractional errors 
similar to those reported in the spurious BICEP2 detection \cite{Ade:2014xna}. 
Of course no massaging of poorly resolved data is going to extract a 
precision bound, but the scatter test seems clearly better.

Note that it is simple to adapt the scatter test to data fits. For example, 
the usual parameterization of the scalar data \cite{Hinshaw:2012aka,
Hou:2012xq,Sievers:2013ica,Ade:2013zuv} implies,
\begin{equation}
\Delta^2_{\mathcal{R}}(k) \simeq A_s \Bigl( \frac{k}{k_0}\Bigr)^{n_s - 1}
\Longrightarrow \quad \Delta^2_{h}(k) \simeq \Delta^2_{h}(k_*)
\Biggl[ 1 + \frac{r(k_*)}{8 (1 \!-\! n_s)} \Bigl[ \Bigl(\frac{k}{k_*}
\Bigr)^{1-n_s} \!-\! 1\Bigr]\Biggr]^{-1} \!. \label{predict}
\end{equation}
Here $A_s$ is the scalar amplitude, $n_s$ is the scalar spectral index,
and $k_0$ is a fiducial wave number.

Finally, we can look forward to the day, in the far future, when the 
tensor power spectrum is well resolved. Then the sort of scatter test
we propose could be employed to search for correlations between features
in the two power spectra. For example, Figure \ref{approximation} depicts 
the bump in the first slow roll parameter  from a model \cite{Adams:2001vc,
Mortonson:2009qv} introduced to explain the scalar power spectrum's dip 
at $\ell \approx 22$ and peak at $\ell \approx 40$ \cite{Covi:2006ci,
Hamann:2007pa,Hazra:2014goa,Hazra:2016fkm}. These features are caused
by the way the scalar nonlocal corrections (\ref{scalarnonlocal}) depend 
upon derivatives of $\epsilon(n)$. The tensor nonlocal corrections 
(\ref{tensornonlocal}) involve the same derivatives --- although lacking 
the large factors of $1/\epsilon$ --- so it is obvious there will be 
corresponding features \cite{Brooker:2017kjd}. Resolving this sort of 
correlation probes the functional relation between the two power spectra 
far more deeply than the single scalar consistency relation. 

\centerline{\bf Acknowledgements}

This work was partially supported by the European Union's Seventh 
Framework Programme (FP7-REGPOT-2012-2013-1) under grant agreement 
number 316165; by the European Union's Horizon 2020 Programme
under grant agreement 669288-SM-GRAV-ERC-2014-ADG;
by a travel grant from the University of Florida International Center, 
College of Liberal Arts and Sciences, Graduate School and Office of 
the Provost; by NSF grant PHY-1506513; and by the Institute for 
Fundamental Theory at the University of Florida. One of us (NCT) 
would like to thank AEI at the University of Bern for its hospitality
while this work was partially completed.

\end{document}